\documentclass[%%final            % use final for the camera ready runs  %%FRK
draft            % use draft while you are working on the paper          %%FRK
,numberedheadings]{aipproc}

\layoutstyle{6x9}

\usepackage{latexsym,amssymb,amsmath,amsfonts,epsfig,revsymb}

\newcommand {\version}{v3}

\newcommand{\vk}{\mathbf{k}}

\newcommand{\dd}{\mathrm{d}}                    % differential d
\newcommand{\id}{\mathrm{d}}                    % integral measure d
\newcommand{\ii}{\mathrm{i}}                    % imaginary i
\newcommand{\diag}{\ensuremath{\mathrm{diag}}}  % diagonal matrix
\newcommand{\modMaxth}{modified-Maxwell theory} %

\def\d{\mathrm{d}}
\newcommand{\st}{spacetime}
\newcommand{\stf}{spacetime foam}

\newcommand{\half}{{\textstyle \frac{1}{2}}}
\newcommand{\bR}{\mathbb{R}}

\newcommand{\dx}{\!\mathsf{d}^4x\,\,}

\newcommand{\bdi}{\begin{displaymath}}
\newcommand{\edi}{\end{displaymath}}
\newcommand{\bfi}{\begin{figure}}
\newcommand{\efi}{\end{figure}}

\newcommand{\beq}{\begin{equation}} %%set eq numbers
\newcommand{\eeq}{\end{equation}}
\newcommand{\beqa}{\begin{eqnarray}}
\newcommand{\eeqa}{\end{eqnarray}}

\newcommand {\bcs}    {boundary conditions}

\newcommand {\YMth}   {Yang--Mills theory}

\newcommand {\lP}{l_\mathsf{Planck}}

\newcommand {\gsim}{\mathrel{\hbox{\rlap{\lower.55ex \hbox {$\sim$}}
            \kern-.3em \raise.4ex \hbox{$>$}}}}
\newcommand {\lsim}{\mathrel{\hbox{\rlap{\lower.55ex \hbox {$\sim$}}
            \kern-.3em \raise.4ex \hbox{$<$}}}}
\begin{document}
%\date{\today\;\;(\version)} %%FRK-draft
\date{}                     %%FRK-preprint

\title[Small-Scale Structure of Spacetime]
      {Small-Scale Structure of Spacetime:\\Bounds and Conjectures}
\title[\mbox{arXiv:0710.3075\;\,[hep-ph]\;\,(\version)}]
      {Small-Scale Structure of Spacetime:\\Bounds and Conjectures}%%FRK-preprint


\author{F.R.~Klinkhamer}{
  address={Institute for Theoretical Physics,
           University of Karlsruhe (TH), 76128 Karlsruhe, Germany}
  %%,email={frans.klinkhamer@physik.uni-karlsruhe.de}
}

\begin{abstract}
This review consists of two parts. The first part establishes certain
astrophysical bounds on the smoothness of classical spacetime.
Some of the best bounds to date
are based on the absence of vacuum Cherenkov radiation
in ultrahigh-energy cosmic rays.
The second part discusses possible implications of these bounds
for the quantum structure of  spacetime.
One conjecture is that the fundamental length scale
of quantum spacetime may be different from the Planck length.
\end{abstract}

\maketitle

\section{INTRODUCTION}
\label{sec:INTRODUCTION}

The present contribution addresses the following basic question:
\emph{does space remain
smooth as one probes smaller and smaller distances?}
A conservative limit on the typical
length scale $\ell$ of any nontrivial small-scale structure of space
results from the fact that the experimental data from
particle accelerators such as LEP at CERN and the Tevatron at Fermilab
are perfectly well described by relativistic quantum fields over
a smooth manifold (specifically, Minkowski spacetime),
\beq
\ell\big|^\mathsf{LEP/Tevatron}  \lesssim 10^{-18}\:\mathsf{m}
\approx \hbar\, c/ \big( 200\,\mathsf{GeV} \big) \,.
\label{eq:LEP-bound}
\eeq
Remark that
the last approximate equality in \eqref{eq:LEP-bound} follows from the
Heisenberg uncertainty principle applied to the matter probes, even though
the bounded spacetime length $\ell$ itself may be a purely classical quantity.

Yet, astrophysics provides us with very much higher energies to probe \st.
A possible strategy is then
\begin{itemize}
\item
to consider the phenomenology of simple spacetime models;
\item
to obtain bounds on the model parameters
from ultrahigh-energy cosmic rays;
\item
to establish the main theoretical implications.
\end{itemize}

Some of our recent work has pursued the above strategy and the aim
of the present article is to review this work in a coherent fashion.
In Sec.~\ref{sec:PHENOMENOLOGY}, we discuss the
phenomenology of two simple models for modified photon propagation
\cite{BernadotteKlinkhamer2007,KaufholdKlinkhamer2007}.
In Secs.~\ref{sec:UHECRboundsModel1} and \ref{sec:UHECRboundsModel2},
we obtain ultrahigh-energy cosmic-ray (UHECR) bounds
\cite{BernadotteKlinkhamer2007,KaufholdKlinkhamer2007,KlinkhamerRisse2007}
on the parameters of the two types of models considered
and, in Sec.~\ref{sec:UHECRboundsImplications}, we discuss some implications.
In Sec.~\ref{sec:CONJECTURES}, which can be read independently of
Secs.~\ref{sec:PHENOMENOLOGY} and \ref{sec:UHECR-CHERENKOV-BOUNDS},
we put forward two conjectures  \cite{Klinkhamer2007} on the fundamental
length of a hypothetical small-scale structure of quantum spacetime.
In Sec.~\ref{sec:SUMMARY}, we summarize our results.

%%\newpage%%FRK
\section{PHENOMENOLOGY}
\label{sec:PHENOMENOLOGY}

Two simple models of photon propagation will be
presented in this section.
The first model has the standard quadratic Maxwell action
density term integrated over a flat \st~manifold with
``defects.'' For this first model, the modified photon
dispersion relation will be calculated in the long-wavelength
approximation.

The second model has a quadratic  modified-Maxwell
term integrated over standard Minkowski spacetime.
The corresponding photon dispersion relation can be readily obtained.
Having standard Dirac fermions coupled to photons with
modified propagation properties, the  process of
``vacuum Cherenkov radiation'' may be allowed for certain combinations
of Lorentz-violating parameters. For the second model,
the relevant Cherenkov energy threshold will be given explicitly.

\subsection{Model 1}
\label{sec:PhenomenologyModel1}

Consider standard Quantum Electrodynamics (QED) \cite{ItzyksonZuber1980}
over a classical spacetime-foam\footnote{A note of caution may
be in order, as the terminology ``spacetime foam'' is often considered
to refer solely to the quantum structure of spacetime
\cite{Wheeler1957,Wheeler1968,Hawking1978,Visser1996},
whereas, here, the picturesque designation ``classical spacetime-foam manifold''
simply refers to a classical manifold with nontrivial small-scale
structure (resembling, for example, a well-known Swiss cheese).
The elusive ``quantum structure of spacetime'' will be discussed
further in Sec.~\ref{sec:CONJECTURES}, whereas we remain with the
more or less familiar classical spacetime
in Secs.~\ref{sec:PHENOMENOLOGY} and \ref{sec:UHECR-CHERENKOV-BOUNDS}.}
 manifold (with details to be specified later), which has the following action:
\beq
S_\mathsf{Model\,1} \!=\!
\int_\mathsf{foam}^{\,\prime} \dx
  \Bigl(
  -{\textstyle\frac{1}{4}}\, \eta^{\mu\rho}\eta^{\nu\sigma}
  \,F_{\mu\nu}(x)\,F_{\rho\sigma}(x)
+
\overline\psi(x) \Big( \gamma^\mu
\big(\ii\,\partial_\mu -e\, A_\mu(x) \big) -M\Big) \psi(x)
  \Bigr),
\label{eq:model1-action}
\eeq
where $F_{\mu\nu}(x)\equiv \partial_\mu
A_\nu(x)-\partial_\nu A_\mu(x)$ is the Maxwell field strength tensor
of the gauge
field $A_\mu(x)$ and $\psi(x)$ the Dirac spinor field corresponding to,
e.g., a proton with electric charge $e$ and rest mass $M$.

\begin{figure}[t]\label{fig:foam}
\includegraphics[height=.2\textheight]{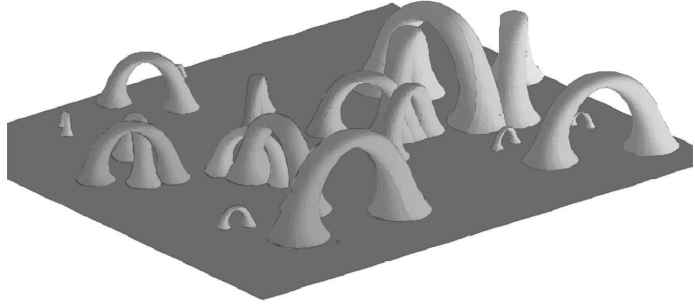}
\caption{Example of a static (time-independent) classical \st~foam,
with one spatial dimension suppressed. The corresponding spacetime is mostly flat
(Minkowski-like) and has several permanent wormholes added.}
\end{figure}

\begin{figure}[b]\label{fig:defect}
\includegraphics[height=.175\textheight]{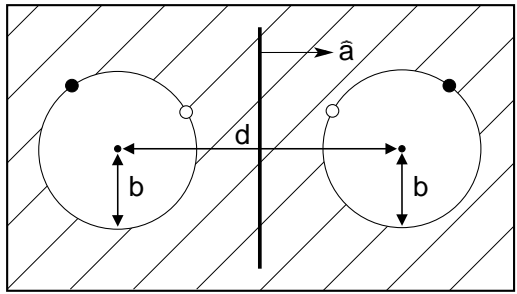}
\caption{Three-space from a single
wormhole-like defect (two spheres with equal
radius $b$ and distance $d > 2b$ between their centers) embedded
in $\bR^3$, with the ``interiors'' of the two spheres removed
and their points identified by reflection in a central plane with
normal unit vector $\widehat{\mathbf{a}}$.
If the ``long distance'' $d$ between the centers of the wormhole
mouths is of the same order of magnitude as their width $2\,b$,
the defect is effectively localized in the ambient three-space.}
\end{figure}

The prime on the integral sign of \eqref{eq:model1-action}
indicates the restriction to the flat Minkowski part
of the manifold with specific \bcs~at certain submanifolds
called ``defects.'' The procedure can perhaps best be illustrated
by the example of Fig.~\ref{fig:foam} on the next page, where the
static ``wormholes'' \cite{Wheeler1957,Wheeler1968}
are sliced off (or, more accurately, the wormhole ``throats'' are taken
to have zero lengths) and the resulting holes/defects
in flat spacetime are given appropriate \bcs~for the
vector and spinor fields appearing in \eqref{eq:model1-action};
see Fig.~\ref{fig:defect} for a sketch.

The coordinates of flat spacetime are denoted by  $(x^\mu)$ $=$
$(x^0$, $\mathbf{x})$ $=$ $(c\,t$, $x^1$, $x^2$, $x^3)$
and the standard Minkowski metric is $(\eta_{\mu\nu})$ $=$
$\diag(1$,$-1$,$-1$,$-1)$. The %%totally antisymmetric
Levi-Civita symbol $\epsilon_{\mu\nu\rho\sigma}$, which will be used later,
is normalized by $\epsilon_{0123}=1$.
The direction of a 3--vector $\mathbf{x}$ is given by the unit
3--vector $\widehat{\mathbf{x}}\equiv\mathbf{x}/|\mathbf{x}|$.
In this section and the next, we use natural units with $c=\hbar=1$,
but, occasionally, we display $c$ or $\hbar$ in order to clarify
the physical dimensions of a particular expression. The dimensionful
constants $c$ and $\hbar$  will, however, occupy the center of the stage
in Sec.~\ref{sec:CONJECTURES}.

Next, take some very simple classical \st-foam models with:
\begin{itemize}
\item
identical defects (spacetime size $\overline{b}$) embedded
in Minkowski spacetime,
\item
a homogeneous and isotropic distribution of defects
$\big($spacetime density $n \equiv 1/\overline{l}^{\,4}\,\big)$,
\item
a strong dilution of defects ($\overline{b} \ll \overline{l}$),
\end{itemize}
where the last requirement, in particular, is a purely
technical assumption in order to simplify the calculation.

Remark that, for the classical \st~shown in Fig.~\ref{fig:foam},
there is no topology change (in fact, it is unclear whether or not topology
change is allowed at all; see, e.g., Ref.~\cite{Visser1996}).
As far as our calculations are concerned, the detailed dynamics
of the classical defects considered does not appear to be important;
what matters are average quantities such as the typical defect size and
separation.

Now calculate the proton and photon dispersion relations
in the long-wavelength
approximation, $\lambda \gg \overline{l} \gg \overline{b}$.
No details of the calculation \cite{BernadotteKlinkhamer2007}
will be given here but only the heuristics for the photon case:
\emph{localized defects correspond to fictional multipoles}
which affect the propagation of electromagnetic plane waves.
Electromagnetic-wave propagation over a
spacetime with defects is then different from
propagation over a perfectly smooth Minkowski \st.
In fact, the basic physics is the same as that of the so-called
``Bethe holes'' in waveguides \cite{Bethe1944}.
It is clear that different types of localized defects
give essentially the same result,
just with different numerical coefficients.

The modified proton ($p$) and photon ($\gamma$) dispersion relations
for wave number $k \equiv |\vk| \equiv 2\pi/\lambda$
can, therefore, be given in completely general form
(that is,  independent of the detailed calculations):
\begin{subequations}\label{eq:disprel-general-form}
\beqa \omega^{2}_{p} &\equiv&
\widetilde{m}_{p}^2\,c_{p}^4/\hbar^2+ c_{p}^2\, k^2
+\mathsf{O}\big(c_{p}^2\,\widetilde{b}^2\, k^4\big)\,,
\label{eq:disprel-general-form-proton}
\\[1mm]
\omega^{2}_\gamma &=&
\big(1+ \widetilde{\sigma}_2\,\widetilde{F}\,\big)\; c_{p}^2\, k^2 +
\widetilde{\sigma}_4\;\widetilde{F}\;c_{p}^2\,\widetilde{b}^2\, k^4
+ \mathsf{O}\big(c_{p}^2\,\widetilde{b}^4\, k^6\big)\,,
\label{eq:disprel-general-form-photon}
\eeqa
\end{subequations}
with $c_{p}^2$ simply defined as the coefficient of
the quadratic proton term, effective on/off factors
$\;\widetilde{\sigma}_2, \widetilde{\sigma}_4 \in \{-1,0,+1\}$,
an effective defect size $\widetilde{b}$,
and an effective excluded-volume factor
\beq\label{eq:widetildeF}
\widetilde{F} \equiv \big(\,\widetilde{b}/\widetilde{l}\;\big)^4,
\eeq
which, for the moment, is assumed to be much less than unity.

Observe that the calculated dispersion relations as given by
(\ref{eq:disprel-general-form}ab)
do not contain cubic terms in $k$, consistent with general
arguments based on coordinate independence and
rotational invariance \cite{Lehnert2003}.
Furthermore, the photon dispersion relations found are the same for
both polarization modes (absence of birefringence) because
of the assumed isotropy of the defect distributions.
However, the photon dispersion relations do show birefringence,
but still no cubic terms, if there is assumed to be
an anisotropic distribution of intrinsically asymmetric defects
(having, for example, defects from
Fig.~\ref{fig:defect} with all axes $\widehat{\mathbf{a}}$ aligned).

As mentioned above, calculations of certain simple spacetime-foam models
give the effective dispersion-relation parameters (with tildes)
in terms of the underlying spacetime parameters (with bars):
\beq\label{eq:parametersfromtau1}
 \widetilde{b}
=\beta\; \overline{b}   \,,\;\; \widetilde{l} =\chi\; \overline{l}
\,,\;\; \widetilde{\sigma}_2 =-1\,,\;\; \widetilde{\sigma}_4 =1\;, \eeq
where $\beta$ and $\chi$ are positive numerical constants of order unity,
which depend on the details of the defects considered.
Note that, for static defects as
in Fig.~\ref{fig:defect}, with size $b$ and separation $l$,
a physically more appropriate definition of $\widetilde{F}$
would be $(b/l)^3$, but, mathematically,
definition \eqref{eq:widetildeF} with $(b/l)^4$ can still be used.

With intrinsically different \bcs~at the defect
locations for the different types of matter fields (spinor and vector),
the quadratic terms of the proton and photon
dispersion relations can be expected to be different in general.
Having defined the quadratic proton term
in \eqref{eq:disprel-general-form-proton} as $c_{p}^2\, k^2$,
the quadratic photon term in \eqref{eq:disprel-general-form-photon}
will then differ from $c_{p}^2\, k^2$,
unless the defects are infinitely small or infinitely far apart.
The resulting unequal proton and photon velocities may lead to
new types of decay processes \cite{KaufholdKlinkhamer2006},
for example, vacuum Cherenkov radiation $p \to p\,\gamma$ for the case
of a negative coefficient $\widetilde{\sigma}_2\,\widetilde{F}$
in \eqref{eq:disprel-general-form-photon}. This particular decay
process will be discussed further in the next subsection.

Regardless of the origin and interpretation, it is clearly important to
obtain bounds on the effective parameters $\widetilde{F}$
and $\widetilde{b}$ in the modified
photon dispersion relation \eqref{eq:disprel-general-form-photon}
for a proton dispersion relation
given by \eqref{eq:disprel-general-form-proton}.

%%\newpage%%FRK
\subsection{Model 2}
\label{sec:PhenomenologyModel2}

In the previous subsection, we have found a
quadratic photon term of the dispersion relation which was modified
by a single parameter $\widetilde{F}$ determined by the
underlying structure of spacetime. This type of modified photon propagation
can easily be generalized.

Consider the following action for a Lorentz-violating deformation of QED:
\beq\label{eq:model2-action}
 S_\mathsf{Model\,2} =
S_\mathsf{modM}+S_\mathsf{standD}\,,
\eeq
with a modified-Maxwell term
\cite{ChadhaNielsen1982,ColladayKostelecky1998},
\beq\label{eq:modM-action}
S_\mathsf{modM} = \int_{\mathbb{R}^4} \id^4 x \;
\Big( -\textstyle{\frac{1}{4}}\, \big( \eta^{\mu\rho}\eta^{\nu\sigma}
+\kappa^{\mu\nu\rho\sigma}\big) \, F_{\mu\nu}(x)\,F_{\rho\sigma}(x)
\Big)\,, \eeq
and the standard Dirac term \cite{ItzyksonZuber1980} for a spin--$\half$ particle
with charge $e$ and mass $M$,
\beq\label{eq:modMstandD-action}
S_\mathsf{standD} =\int_{\mathbb{R}^4} \id^4 x \; \overline\psi(x) \Big(
\gamma^\mu \big(\ii\,\partial_\mu -e\, A_\mu(x) \big) -M\Big) \psi(x)\,.
\eeq
Theory \eqref{eq:model2-action} is gauge-invariant,
CPT--even, and power-counting renormalizable.

The quantity $\kappa^{\mu\nu\rho\sigma}$
in the modified-Maxwell term \eqref{eq:modM-action}
is a constant background tensor with real and dimensionless components.
This background tensor $\kappa^{\mu\nu\rho\sigma}$ has, in fact, the same
 symmetries as the Riemann curvature tensor and a double trace condition
$\kappa^{\mu\nu}_{\phantom{\mu\nu}\mu\nu}=0$, so that there are $20-1=19$
independent components. All components of
the $\kappa$--tensor in \eqref{eq:modM-action} are
assumed to be sufficiently small in order to ensure energy positivity.

As the ten birefringent parameters are already constrained
at the $10^{-32}$ level
\cite{KosteleckyMewes2002}, restrict the theory to the nonbirefringent
sector by making the following \emph{Ansatz} \cite{BaileyKostelecky2004}:
\beq\label{eq:nonbirefringent-ansatz}
\kappa^{\mu\nu\rho\sigma} = \textstyle{\frac{1}{2}}
\big(\,
 \eta^{\mu\rho}\,\widetilde{\kappa}^{\nu\sigma}
-\eta^{\nu\rho}\,\widetilde{\kappa}^{\mu\sigma}
+\eta^{\nu\sigma}\,\widetilde{\kappa}^{\mu\rho}
-\eta^{\mu\sigma}\,\widetilde{\kappa}^{\nu\rho}
\,\big) , \eeq
for a
symmetric and traceless matrix $\widetilde{\kappa}^{\mu\nu}$ with $10-1=9$
independent components.
Rewrite these nine Lorentz-violating ``deformation parameters''
$\widetilde{\kappa}^{\mu\nu}$ as follows:
\newcommand{\third}{\textstyle{\frac{1}{3}}}
\beq\label{eq:widetilde-kappa-mu-nu-Ansatz}
\big(\widetilde{\kappa}^{\mu\nu}\big) \equiv
\mathsf{diag}\big(1,\third,\third,\third\big)\, \overline{\kappa}^{00}
+\big(\delta\widetilde{\kappa}^{\mu\nu}\big),\;\;
\delta\widetilde{\kappa}^{00}=0\,, \eeq
with one independent parameter
$\overline{\kappa}^{00}$ for the spatially isotropic part of
$\widetilde{\kappa}^{\mu\nu}$ and eight independent parameters
$\delta\widetilde{\kappa}^{\mu\nu}$ which need not be smaller than
$\overline{\kappa}^{00}$.

For later use, also define a vector
$\vec{\alpha}$ in parameter space $\mathbb{R}^9$:
\beq\label{eq:alpha-parameters}
\vec{\alpha}
\equiv \left(
  \begin{array}{c}
    \alpha^0 \\
    \alpha^1 \\
    \alpha^2 \\
    \alpha^3 \\
    \alpha^4 \\
    \alpha^5 \\
    \alpha^6 \\
    \alpha^7 \\
    \alpha^8 \\
  \end{array}
\right)
 \equiv
\left(
  \begin{array}{c}
    \widetilde{\alpha}^{00} \\
    \widetilde{\alpha}^{01} \\
    \widetilde{\alpha}^{02} \\
    \widetilde{\alpha}^{03} \\
    \widetilde{\alpha}^{11} \\
    \widetilde{\alpha}^{12} \\
    \widetilde{\alpha}^{13} \\
    \widetilde{\alpha}^{22} \\
    \widetilde{\alpha}^{23} \\
  \end{array}
\right) \equiv \left(
  \begin{array}{c}
    (4/3)\,\overline{\kappa}^{00}\\
    2\,\delta\widetilde{\kappa}^{01} \\
    2\,\delta\widetilde{\kappa}^{02} \\
    2\,\delta\widetilde{\kappa}^{03} \\
    \delta\widetilde{\kappa}^{11} \\
    \delta\widetilde{\kappa}^{12} \\
    \delta\widetilde{\kappa}^{13} \\
    \delta\widetilde{\kappa}^{22} \\
    \delta\widetilde{\kappa}^{23} \\
  \end{array}
\right), \eeq
which is taken to have the standard Euclidean norm,
\begin{equation}\label{eq:alpha-space-norm}
|\vec{\alpha}|^2 \equiv \sum_{l=0}^{8}\:\big(\alpha^l\,\big)^2\,.
\end{equation}
Note that the negative of the isotropic parameter $\alpha^0$ corresponds,
in leading order, to the coefficient $\widetilde{\sigma}_2\,\widetilde{F}$
of the modified photon dispersion relation \eqref{eq:disprel-general-form-photon}
calculated for Model 1.

It may be of interest to mention that, whereas Model 2 has
constant deformation parameters \eqref{eq:nonbirefringent-ansatz},
another model has a
small stochastic parameter $g(x)$
multiplying a CPT--odd action density term
$\epsilon^{\mu\nu\rho\sigma}\,F_{\mu\nu}(x)\,F_{\rho\sigma}(x)$.
This particular stochastic model
has been studied in the long-wavelength approximation in
Ref.~\cite{KlinkhamerRupp2004}. Recently, we have  become aware
of a complementary paper \cite{HuShiokawa1998}, which considers
similar models in the short-wavelength approximation.

\begin{figure}\label{fig:diagram}
  \includegraphics[height=.125\textheight]{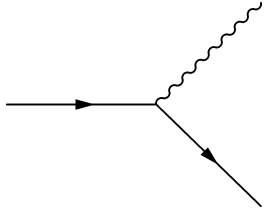}
  \caption{Feynman diagram for vacuum Cherenkov radiation.}
\end{figure}

Model 2 has, for appropriate deformation parameters $\vec{\alpha}$,
a maximum proton velocity larger than the phase velocity of light, which
allows for vacuum Cherenkov radiation $p \to p\,\gamma$.
This particular decay process has
been studied classically by Altschul \cite{Altschul2007PRL98}
and quantum-mechanically at tree-level (Fig.~\ref{fig:diagram})
by Kaufhold and the present
author \cite{KaufholdKlinkhamer2007,KaufholdKlinkhamer2006}.
The radiated-energy rate of a primary (on-shell) particle with
point charge $Z_\mathsf{prim}\,e$, mass $M_\mathsf{prim}> 0$,
momentum $\mathbf{q}_\mathsf{prim}$, and ultrarelativistic energy
$E_\mathsf{prim} \sim c\,|\mathbf{q}_\mathsf{prim}|$
is asymptotically given by \cite{KaufholdKlinkhamer2007}
\beq\label{eq:dWdt-asymptotic}
\frac{\dd W_\mathsf{Model\,2}(\widehat{\mathbf{q}}_\mathsf{prim},\,E_\mathsf{prim})}
     {\dd t}\,
\Bigg|_{E_\mathsf{prim}^2 \gg E^2_\mathsf{thresh}}\!
\sim \;\;
Z_\mathsf{prim}^2\; \frac{e^2}{4\pi}\;
\xi(\widehat{\mathbf{q}}_\mathsf{prim})\; E_\mathsf{prim}^2 / \hbar\;, \eeq
with a nonnegative dimensionless
coefficient $\xi(\widehat{\mathbf{q}}_\mathsf{prim})$
from appropriate contractions of the $\kappa$--tensor with two
rescaled $q$--vectors and the squared threshold energy \cite{Altschul2007PRL98}
\beq\label{eq:Ethreshold}
 E_\mathsf{thresh}^2 =
\frac{M_\mathsf{prim}^2\, c^4}{R_\epsilon\big(\,\alpha^{0}+
\alpha^{j}\;\widehat{\mathbf{q}}_\mathsf{prim}^{j} +\widetilde{\alpha}^{jk}\;
\widehat{\mathbf{q}}_\mathsf{prim}^{j}\,\widehat{\mathbf{q}}_\mathsf{prim}^{k}\,\big)}
+ \mathsf{O}\big(M_\mathsf{prim}^2\, c^4\big),
\eeq
in terms of a regularized ramp function
$R_\epsilon(x) \equiv \epsilon+ (x + |x|\,)/2$ for a real
variable $x$ and an arbitrarily small positive parameter $\epsilon$.

Exact tree-level results have been obtained
recently \cite{KaufholdKlinkhamerSchreck2007}
for the restricted model (labeled ``isotropic case'')
with $\alpha_{0}\equiv\alpha^{0}>0$ and $\alpha^l=0$, for $l=1,\ldots,8$.
Setting again $\hbar=c=1$, the radiated-energy rate for a
spin--$\half$ Dirac particle (charge $e$, mass $M$, and energy $E$
above threshold) is given by the following expression:
\beqa\label{eq:dWdt-exact-isotropic}
\hspace*{-10mm}
\frac{\dd W_\mathsf{Model\,2}^\mathsf{isotropic\:case}}{\dd t}\,
&=&\frac{e^2}{4\pi}\;\,\frac{1}{3\,\alpha_0^3 \,E\,\sqrt{E^2-M^2}}\;
\left(\sqrt{\frac{2-\alpha_0}{2+\alpha_0}}\:E-\sqrt{E^2-M^2}\right)^2
\nonumber\\
\hspace*{-10mm}
&&\times \Bigg\{2\Big(\alpha_0^2+4\alpha_0+6\Big)\,E^2-\Big(2+\alpha_0\Big)
\nonumber\\
\hspace*{-10mm}
&&\times \left(3\,\big(1+\alpha_0\big)\,M^2+2\,\big(3+2\alpha_0\big)\,
\sqrt{\frac{2-\alpha_0}{2+\alpha_0}}\:E\,\sqrt{E^2-M^2}\right)\Bigg\}\,.
\eeqa
The high-energy expansion of \eqref{eq:dWdt-exact-isotropic}
for fixed parameters $\alpha_0$ and $M$ reads
\beqa\label{eq:dWdt-asymptotic-isotropic}
\hspace*{-4mm}
\frac{\dd W_\mathsf{Model\,2}^\mathsf{isotropic\:case}}{\dd t}\,
&=&\frac{e^2}{4\pi}\;\Bigg\{
\left(\frac{7}{24}\,\alpha_0-\frac{1}{16}\,\alpha_0^2+\mathsf{O}(\alpha_0^3)\right)
\,E^2
\nonumber\\
\hspace*{-4mm}
&&+ \left(-1+\frac{1}{48}\,\alpha_0-\frac{3}{32}\,\alpha_0^2+
\mathsf{O}(\alpha_0^3)\right)M^2+
\mathsf{O}\left(\frac{M^4}{\alpha_0\,E^2}\right)\Bigg\}\,,
\eeqa
which shows the quadratic behavior of \eqref{eq:dWdt-asymptotic} with
factor $Z_\mathsf{prim}^2=1$ and constant coefficient $\xi=(7/24)\,\alpha_0$.
From \eqref{eq:dWdt-exact-isotropic}, one also obtains the exact
threshold energy (temporarily reinstating $c$):
\beq\label{eq:Ethreshold-isotropic}
E^\mathsf{Model\,2,\,isotropic\:case}_\mathsf{thresh} =
\frac{M c^2}{\sqrt{\alpha_0}}\; \sqrt{1+\alpha_0/2}\:,
\eeq
which reproduces \eqref{eq:Ethreshold} for $M=M_\mathsf{prim}$
and small enough positive $\alpha_{0}\equiv\alpha^{0}$.
Incidentally, the numerical
value of $\alpha_{0}$ cannot be too large, as the factors
$\sqrt{2-\alpha_0}$ in \eqref{eq:dWdt-exact-isotropic} make clear.

Returning to Model 2 with generic deformation parameters $\alpha^l$,
it is to be expected that  vacuum Cherenkov radiation
only occurs for those parameters $\vec{\alpha}$ for which the
phase velocity of light is less than the maximal attainable
velocity $c$ of the charged particle in theory \eqref{eq:model2-action},
$v_\mathsf{ph} < c$. In fact, this phase-velocity condition
corresponds to having a positive argument of the ramp function
on the right-hand side of \eqref{eq:Ethreshold}.
The relevant domain in parameter space is given by
\beq\label{eq:Dcausalopen}
D_\mathsf{causal}^\mathsf{(open)} \equiv \big\{ \vec{\alpha} \in \mathbb{R}^9
: \; \forall_{\widehat{\mathbf{x}}\in \mathbb{R}^3} \;
\big(\alpha^0 +  \alpha^j\;\widehat{x}^j+
\widetilde{\alpha}^{jk}\;\widehat{x}^j\,\widehat{x}^k \big) >0 \,\big\}\,,
\eeq
for arbitrary unit 3--vector $\widehat{\mathbf{x}}\equiv\mathbf{x}/|\mathbf{x}|$.
The superscript `(open)' in \eqref{eq:Dcausalopen} refers to the use of
the open relation symbol `$>$'  on the right-hand side instead of
the closed symbol  `$\geq$',
because there is no vacuum Cherenkov radiation for the case of $v_\mathsf{ph}=c$.
Most likely, the domain
\eqref{eq:Dcausalopen} constitutes a significant part of the physical
domain of theory \eqref{eq:model2-action}, where, e.g., microcausality holds.
Hence, the subscript `causal' in \eqref{eq:Dcausalopen}.

Microcausality of Lorentz-violating theories has been discussed in, for example,
Refs.~\cite{AdamKlinkhamer2001,KosteleckyLehnert2001,BrosEpstein2002}.
To this author, it is clear that violation of microcausality is
unacceptable in the type of theories considered here, because,
for a spacelike separation of two events, the time order
can be interchanged by an appropriate observer Lorentz transformation.
Perhaps not unacceptable (at least, according to our current experimental
knowledge) is a new class of particle instabilities which occur at ultralarge
three-momentum due to energy non-positivity (at least, in ``concordant''
frames, where the Lorentz-violating parameters are relatively small).

Let us end this subsection with a general remark on Model 2
(see also Sec.~\ref{sec:UHECRboundsImplications} for further discussion).
New phenomena at the energy scale
$E_\mathsf{Planck}\approx 10^{19}\,\mathsf{GeV}$
\cite{Wheeler1957,Wheeler1968} may lead to Lorentz violation
in the low-energy theory, partially described by the model
action  \eqref{eq:model2-action}.
But the resulting Lorentz-violating parameters $\vec{\alpha}$ need
not be extremely small (e.g., sup\-pressed by in\-ver\-se
powers of $E_\mathsf{Planck}$) and can be of order unity, as
long as the theory remains physically consistent.
In fact, the modified-dispersion-relation calculations from
the previous subsection provide an example: the quadratic
coefficient $\widetilde{F}$ in \eqref{eq:disprel-general-form-photon}
can, in principle, be close to one.
Hence, it is important to obtain as strong bounds
as possible on \emph{all} deformation parameters $\vec{\alpha}$.

%%\newpage%%FRK
\section{UHECR CHERENKOV BOUNDS}
\label{sec:UHECR-CHERENKOV-BOUNDS}

The goal of this section is to establish bounds on the parameters
of the two Lorentz-violating models discussed in the previous section.
One type of bounds relies on the process of
vacuum Cherenkov radiation (Fig.~\ref{fig:diagram})
already mentioned in Sec.~\ref{sec:PhenomenologyModel2}.

The basic idea is simple \cite{Beall1970,ColemanGlashow1997}:
\begin{itemize}
\item
if vacuum Cherenkov radiation has a threshold energy
$\!E_\mathsf{thresh}(\widetilde{b},\widetilde{l},\vec{\alpha})$
and the radiation rate above threshold is not suppressed,
then \mbox{UHECRs} with $E_\mathsf{prim}>E_\mathsf{thresh}$
cannot travel far (certainly not over distances of the order of
megaparsecs), as they rapidly radiate away their energy;
\item
observing an UHECR then implies a primary energy $E_\mathsf{prim}$
at or below threshold,
which gives bounds on combinations of $\widetilde{b}$, $\widetilde{l}$,
and $\vec{\alpha}$.
\end{itemize}
Expanding on the last point,
the following inequality must hold for the \emph{measured} primary energy
$E_\mathsf{prim}$ of an UHECR compared to the \emph{theoretical} result for
the threshold energy:
\beq\label{eq:Cherenkov-condition}
E_\mathsf{prim} \leq
E_\mathsf{thresh}(\widetilde{b},\widetilde{l},\vec{\alpha}),
\eeq
which can be written as an upper bound on the parameters
of the Lorentz-violating theory considered
(here, taken to be $\widetilde{b}$, $\widetilde{l}$, and $\vec{\alpha}$,
from the two models discussed in the two previous subsections).

Incidentally, other types of bounds \cite{BernadotteKlinkhamer2007} have been
obtained
(specifically, from the lack of time dispersion and Rayleigh-like scattering
in a particular $\mathsf{TeV}$ gamma-ray flare from the active galaxy Mkn 421),
but the resulting bounds are, for the moment, less tight
than those from the inferred absence of vacuum Cherenkov radiation.

%%\newpage%%FRK
\subsection{UHECR Bounds -- Model 1}
\label{sec:UHECRboundsModel1}

\begin{figure}[t]\label{fig:UHECRflyseye}
  \includegraphics[height=.175\textheight]{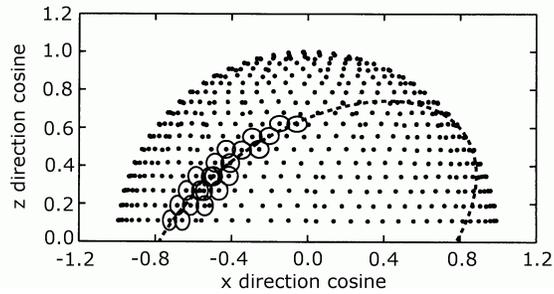}
  \caption{Pointing directions of the 22 photomultiplier tubes which
triggered in connection with the Fly's Eye event \cite{Bird-etal1995} of
October 15, 1991 at 7:34:16 UT.
The pointing directions are shown projected into the $x$--$z$ plane, where
the $x$--axis points east, the $y$--axis north, and the $z$--axis upward.
The triggered phototubes have positive $y$--components.
The dashed line indicates the plane defined
by the shower axis and the detector. See Ref.~\cite{Bird-etal1995}
for further details (figure reproduced by permission of the AAS).
\vspace*{-.5cm}}
\end{figure}

In order to obtain Cherenkov bounds \cite{BernadotteKlinkhamer2007}
for the two parameters $\widetilde{b}$ and $\widetilde{l}$ of
Model 1, start with the spectacular event \cite{Bird-etal1995}
shown in Fig.~\ref{fig:UHECRflyseye} a few pages later.
This particular Extensive Air Shower (EAS) event corresponds, in fact,
to the most energetic particle observed to date, having an energy
$E_\mathsf{prim}\approx 300\,\mathsf{EeV} = 3\times 10^{11}\,\mathsf{GeV}
= 3 \times 10^{20} \, \mathsf{eV}$.

From the EAS event of Fig.~\ref{fig:UHECRflyseye}, we obtain the following
Cherenkov-like bounds on the quadratic and quartic coefficients of
dispersion relation \eqref{eq:disprel-general-form-photon}:
\begin{subequations}\label{eq:Cherenkov-bounds}
\beqa
-3 \times 10^{-23} &\lesssim& \widetilde{\sigma}_2\,
\widetilde{F} \;\;\;\;\; \;\;\;\lesssim\;\;\;
 3 \times 10^{-23}\,,
\label{eq:Cherenkov-bound-ratio}\\[2mm]
 -\left(7 \times 10^{-39}\,\mathsf{m}\right)^{2} &\lesssim&
\widetilde{\sigma}_4\;\widetilde{F}\;\, \widetilde{b}^2
\;\;\lesssim\;\;\;
 \left(5 \times 10^{-38}\,\mathsf{m}\right)^{2}\,,
 \label{eq:Cherenkov-bound-bsquare}
\eeqa
\end{subequations}
based on the detailed analysis of Ref.~\cite{GagnonMoore2004}
(some back-of-the-envelope calculations have been presented
in App.~B of Ref.~\cite{KlinkhamerRupp2005}).
For bounds (\ref{eq:Cherenkov-bounds}ab), the primary was assumed
to be a proton ($p$) with standard partonic distributions and,
as mentioned before, its energy was determined to be
$E_{p} \approx 3 \times 10^{11} \, \mathsf{GeV}$.

Note that the above bounds are two-sided, whereas genuine Cherenkov
radiation would only give one-sided bounds
(the electromagnetic-wave phase velocity must be less than the maximum
particle velocity). The reason for having two-sided bounds is that,
in addition to Cherenkov radiation, another type of process can occur,
namely, proton break-up $p\to p\, e^+\, e^-$
(pair-production by a virtual gauge boson),
which gives the ``other'' sides of the bounds \cite{GagnonMoore2004}.

It is important to understand the dependence of
bounds (\ref{eq:Cherenkov-bounds}ab) on the assumed primary mass
(here, taken as $M_\mathsf{prim}=0.94\;\mathsf{GeV}/c^2$) and primary
energy (here, taken as $E_\mathsf{prim}=3 \times 10^{11} \, \mathsf{GeV}$).
The high and low limits of bounds \eqref{eq:Cherenkov-bound-ratio}
and \eqref{eq:Cherenkov-bound-bsquare} are multiplied by
scaling factors $f_\mathsf{a,\,high/low}$ and
$f_\mathsf{b,\,high/low}$, respectively,  which are
approximately given by
\begin{subequations}\label{eq:Cherenkov-bounds-scaling}
\beqa
f_\mathsf{a,\,high}&\approx&  f_\mathsf{a,\,low} \;\;\;\approx\;\;\;
\big(M_\mathsf{prim}\,c^2/0.94\;\mathsf{GeV}\big)^2\;\;
\big(3 \times 10^{11} \, \mathsf{GeV}/E_\mathsf{prim}\big)^2\,,
\label{eq:Cherenkov-bound-ratio-scaling}
\\[2mm]
f_\mathsf{b,\,high}&\approx& f_\mathsf{b,\,low} \;\;\;\approx\;\;\;
\big(M_\mathsf{prim}\,c^2/0.94\;\mathsf{GeV}\big)^2\;\;
\big(3 \times 10^{11} \, \mathsf{GeV}/E_\mathsf{prim}\big)^4\,.
 \label{eq:Cherenkov-bound-bsquare-scaling}
\eeqa
\end{subequations}
Bounds (\ref{eq:Cherenkov-bounds}ab) would certainly be less
compelling if the Fly's Eye event of Fig.~\ref{fig:UHECRflyseye} were
unique. But, luckily, this is not the case,
as the Pierre Auger Observatory \cite{Abraham-etal2004}
has already seen an event (ID No. $737165$) with a similar energy
of $2\times 10^{11}\,\mathsf{GeV}$ \cite{Abraham-etal2006}.

Let us end this subsection with a somewhat peripheral remark.
The extremely small numbers of order
$10^{-38}\,\mathsf{m} \approx \hbar\, c/ (2 \times 10^{22}\,\mathsf{GeV})$
appearing in the Cherenkov-like bound \eqref{eq:Cherenkov-bound-bsquare},
which traces back to Eq.~(1.14) of Ref.~\cite{GagnonMoore2004},
appear to be incompatible with a quartic-term mass scale
$M_\mathsf{QG2} \approx 6 \times 10^{10}\,\mathsf{GeV}$
claimed \cite{Albert-etal2007-speculation}
to be a possible explanation of time-dispersion effects
observed in the July 9, 2005 gamma-ray flare from Mkn 501
by the MAGIC telescope
(see, in particular, Fig.~7 of Ref.~\cite{Albert-etal2007-exp}).
Even though we have doubts as to the validity of
this possible detection of ``quantum-gravity''
effects,\footnote{Note also that the June 30, 2005 flare,
for the two medium-energy bands
$0.25-0.60\,\mathsf{TeV}$ and $0.60-1.20\,\mathsf{TeV}$
shown in Fig.~6 of Ref.~\cite{Albert-etal2007-exp},
appears to have a time-dispersion  behavior \emph{opposite} to that of
the July 9, 2005 flare, as shown in Fig.~7 of the same reference.}
it is instructive to see how, in principle, astrophysical data could give
more than just bounds on the possible small-scale structure of space.

%%\newpage%%FRK
\subsection{UHECR Bounds -- Model 2}
\label{sec:UHECRboundsModel2}

\begin{figure}[b]\label{fig:Auger}
\includegraphics[height=.25\textheight]{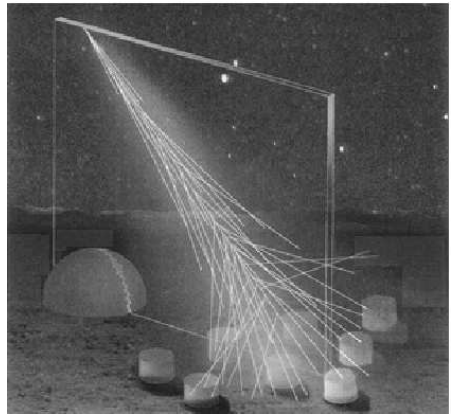}
\caption{The hybrid nature of the Pierre Auger Observatory
\cite{Abraham-etal2004}
provides for two independent ways
to study extensive air showers, namely, by an array of
surface water-Cherenkov detectors  and a collection of
air-fluorescence telescopes [see \protect\url{http://www.auger.org/}].}
\end{figure}

Next, obtain Cherenkov bounds \cite{KaufholdKlinkhamer2007,KlinkhamerRisse2007}
for the nine parameters $\vec{\alpha}$ of Model 2.
Start with the 15 Auger events \cite{Abraham-etal2006}
of Table~\ref{tab-Auger-events-random} on the next page,
where the event identification number is given in column 1,
the energy $E_\mathsf{prim}$ of a hadron primary in column 2,
the shower-maximum atmospheric depth
$X_{\rm max}$ in column 3, and pseudo-random event directions
with right ascension $\mathsf{RA}'\in [0,360^{\circ}]$ and
declination $\mathsf{DEC}'\in [-70^{\circ},25^{\circ}]$ in column 4.\footnote{The
directions of these 15 events
have not yet been released by the Pierre Auger Collaboration and,
for this reason, fictional directions have been given
in  Table~\ref{tab-Auger-events-random} (with primes alerting to
their nonreality).
The bounds of this subsection are based on the numbers given
in Table~\ref{tab-Auger-events-random}, but we are confident that
the same bounds are obtained from the actual event directions
when they are made available by the Pierre Auger Collaboration.
For the bounds obtained here, it only matters that there are 15 UHECR events
with more or less  equal energies $E_\mathsf{prim}\approx 15\:\mathsf{EeV}$
and more or less random directions over a significant part of the
sky, the precise association of energy and direction being irrelevant.}
These are high-quality events, having been observed in the ``hybrid''
mode (Fig.~\ref{fig:Auger}) of the Pierre Auger
Observatory \cite{Abraham-etal2004}.
Specifically, the $X_{\rm max}$ values give
information on the primary particle type, as discussed
in Ref.~\cite{KlinkhamerRisse2007} and references therein.

As mentioned in Sec.~\ref{sec:PhenomenologyModel2}, vacuum Cherenkov
bounds can only be obtained for parameters $\vec{\alpha}$
in domain \eqref{eq:Dcausalopen},
for which the phase velocity of light is less than the maximal attainable
velocity $c$ of the charged particle. Now determine a hypersphere
$S^8_a$ of radius $a$ in this subspace, so that for
\emph{each} $\vec{\alpha} \in S^8_a$ the Cherenkov threshold
condition \eqref{eq:Cherenkov-condition},
with expression \eqref{eq:Ethreshold} inserted, is violated for
\emph{at least one} event from Table~\ref{tab-Auger-events-random}.

The excluded domain of parameter space then corresponds
to the region on or outside this hypersphere $S^8_a$.
Namely, for a positive integrand of the ramp
function on the right-hand side of \eqref{eq:Ethreshold},
the first term on the right-hand side is multiplied by a
factor $1/\lambda$ under scaling of $\vec{\alpha}\to\lambda\,\vec{\alpha}$
(and $\epsilon \to \lambda\,\epsilon$) with $\lambda>1$,
so that inequality \eqref{eq:Cherenkov-condition} is violated for $\lambda>1$
if it is already for $\lambda=1$.
The  events from Table~\ref{tab-Auger-events-random} give at the $2\sigma$ level:
\vspace*{-1mm}%%FRK
\begin{subequations}\label{eq:Dexcluded}
\beqa
D_\mathsf{excluded}^a &=& D_\mathsf{causal}^\mathsf{(open)} \,\cap\,
\big\{\vec{\alpha} \in \mathbb{R}^9 : \; |\vec{\alpha}| \geq a \big\}\,,
\label{eq:Dexcluded-general}\\
a &\approx& 3 \times 10^{-18} \,
\left(\frac{M_\mathsf{prim}}{16\:\mathsf{GeV}/c^2}\right)^{2},
\label{eq:Dexcluded-avalue}
\eeqa
\end{subequations}
where $|\vec{\alpha}|$ is the standard Euclidean norm \eqref{eq:alpha-space-norm}
and where the mass of the primary charged particle has conservatively
been taken equal to that of oxygen (most
primaries in the sample are expected to be protons and helium nuclei).

The resulting
UHECR Cherenkov bounds for nonbirefringent modified-Maxwell theory
can be described as follows:
considering deformation parameters $\alpha^0, \dots, \alpha^8$ with a
corresponding phase velocity of light less than the maximal attainable
velocity of charged particles, each of these nine parameters
must separately have a modulus less than the value $a$ given by
\eqref{eq:Dexcluded-avalue}.
Improving slightly by considering all nine parameters simultaneously,
the new $2\sigma$ bound from astrophysics is given by
\vspace*{-1mm}%%FRK
\beq
\vec{\alpha} \in D_\mathsf{causal}^\mathsf{(open)}:\;
\big|\,\vec{\alpha}\,\big|^2 \equiv
\sum_{l=0}^{8}\:\big(\alpha^l\,\big)^2  < \Big( 3 \times 10^{-18}\Big)^2,
\label{eq:vecalpha-Cherenkov-bound}
\eeq
with $M_\mathsf{prim}$ simply set to $16\:\mathsf{GeV}/c^2$.\newpage
%
% %%FRK--by hand
%
\begin{table}[h]
\begin{tabular}{lccc}
\hline
    \tablehead{1}{l}{b}{ID No.}
  & \tablehead{1}{c}{b}{$\mathbf{E}_\mathbf{prim}$ [EeV]}
  & \tablehead{1}{c}{b}{$\mathbf{X}_\mathbf{max}$
                       [$\mathbf{g}\:\mathbf{cm}^\mathbf{-2}$] }
  & \tablehead{1}{c}{b}{
    (RA$\mathbf{'}$,\,DEC$\mathbf{'}$) [$\mathbf{deg}$]}\\
\hline
 668949$'$    & 18 & 765 & (356 , --29)\\
 673409$'$    & 13 & 760 & (344 , --62)\\
 828057$'$    & 14 & 805 & (086 , --34)\\
 986990$'$    & 16 & 810 & (152 , --33)\\
1109855$'$    & 17 & 819 & (280 , --30)\\
1171225$'$    & 16 & 786 & (309 , --70)\\
1175036$'$    & 18 & 780 & (228 ,  +17)\\
1421093$'$    & 27 & 831 & (079 ,  +13)\\
1535139$'$    & 16 & 768 & (006 , --62)\\
1539432$'$    & 13 & 787 & (153 , --15)\\
1671524$'$    & 14 & 806 & (028 , --63)\\
1683620$'$    & 21 & 824 & (024 , --23)\\
1687849$'$    & 17 & 780 & (031 , --23)\\
2035613$'$    & 12 & 802 & (079 , --08)\\
2036381$'$    & 29 & 782 & (158 , --03)\\
\hline
\vspace*{-4cm}\end{tabular}
\caption{``Hybrid'' events recorded by the Pierre Auger Observatory
\cite{Abraham-etal2004,Abraham-etal2006};
see Sec.~\ref{sec:UHECRboundsModel2} in the main text for further details.}
\label{tab-Auger-events-random}
\end{table}

\newpage%\noindent  %%FRK--by hand

Three remarks are in order. First,
bound \eqref{eq:vecalpha-Cherenkov-bound} holds for the reference
frame in which the cosmic-ray energies are measured.
This frame is essentially given by the rest system of the Sun.
Second, the restriction of $\vec{\alpha}$ to domain \eqref{eq:Dcausalopen}
makes the Cherenkov bound \eqref{eq:vecalpha-Cherenkov-bound}
effectively ``one-sided,'' as discussed in the third paragraph of
Sec.~\ref{sec:UHECRboundsModel1}. To specify the precise domain
in the 9--dimensional parameter space requires, however, some care.
For example, in the 2--dimensional subspace with
only $\alpha^0 \equiv \alpha\, \cos\phi$
and $\alpha^1 \equiv \alpha\, \sin\phi$ nonzero,
the allowed domain from \eqref{eq:vecalpha-Cherenkov-bound}
is given by an open disk segment
with $- \pi/4 < \phi <  \pi/4$ and $0 \leq \alpha < 3 \times 10^{-18}$.
Third, consider the restricted \emph{isotropic} model  with only
parameter $\alpha^0$ nonzero, which was
already discussed in the paragraph containing the exact tree-level
results \eqref{eq:dWdt-exact-isotropic}--\eqref{eq:Ethreshold-isotropic}.
Then, the \emph{single} Fly's Eye event
\cite{Bird-etal1995} with $E_\mathsf{prim}\approx 300\,\mathsf{EeV}$
(or the single $200\,\mathsf{EeV}$ Auger event \cite{Abraham-etal2006}
mentioned in Sec.~\ref{sec:UHECRboundsModel1})
gives a one-sided bound on $\alpha^0$ at the $10^{-23}$ level;
see Eq.~(C6) of Ref.~\cite{KaufholdKlinkhamer2007}.
This single $\alpha^0$ bound corresponds, in fact, to
the negative part of the previous bound \eqref{eq:Cherenkov-bound-ratio}.

At this moment, it may be of interest to recall the best bounds from
laboratory experiments \cite{Saathoff-etal2003,Odom-etal2006,Stanwix-etal2006}:
\begin{itemize}
\item
for the spatially isotropic
nonbirefringent deformation parameter $\alpha^0$,
there is a direct bound at the $10^{-7}$ level
[33(c)] %%FRK--ref by hand
%%\cite[(c)]{Saathoff-etal2003}
and an indirect (electron anomalous-magnetic-moment)
bound at the $10^{-8}$ level
[34(c)];   %%FRK--ref by hand
%%\cite[(c)]{Odom-etal2006};
\item
for the eight other nonbirefringent
parameters $\alpha^1, \dots, \alpha^8$, there are direct bounds at the
$10^{-12}$ to $10^{-16}$ levels \cite{Stanwix-etal2006}.
\end{itemize}
Observe that, for Earth-based cavity experiments
with an orbital velocity $v_\oplus$ around the Sun,
the parity-odd parameters $\alpha^l$, for $l=1,2,3$,
typically enter the measured relative frequency shift
with a factor $(v_\oplus/c) \approx 10^{-4}$, which reduces
the sensitivity for these three parameters.
In these and other experiments,
the parameter $\alpha^0$ even enters with a quadratic boost factor
(precisely which velocity plays a decisive role
depends on the experimental setup) and
the sensitivity is reduced also for this parameter.
See Refs.~\cite{KosteleckyMewes2002,Saathoff-etal2003,Stanwix-etal2006}
for further details.

Astrophysics and laboratory bounds on $\vec{\alpha}$ are not directly
comparable, the latter having all the benefits of repeatability and control.
But the astrophysics bounds at the $10^{-18}$ level are certainly indicative.
Moreover, the UHECR Cherenkov bounds on $\vec{\alpha}$ can be expected to drop
to the $10^{-23}$ level in the coming years \cite{KlinkhamerRisse2007}.

%%\newpage%%FRK
\subsection{UHECR Bounds -- Implications}
\label{sec:UHECRboundsImplications}

In Sec.~\ref{sec:UHECRboundsModel1}, we have obtained
Cherenkov bounds for the effective defect excluded-volume factor
$\widetilde{F}$ and the effective defect size $\widetilde{b}$ in the
modified photon dispersion relation \eqref{eq:disprel-general-form-photon}:
\begin{subequations}\label{eq:Cherenkov-bounds-concl}
\beqa
0 \;\;\leq\;\; \widetilde{F}
&\!\! \lesssim \!\!&  10^{-23}\,,
\label{eq:Cherenkov-bounds-concl-F}\\[1mm]
0 \;\;\leq\;\; \widetilde{b}\;
&\!\! \lesssim \!\!&
 10^{-26} \:\mathsf{m}
 \;\approx\; \hbar\, c/\big( 2 \times 10^{10}\,\mathsf{GeV} \big) \,,
\label{eq:Cherenkov-bounds-concl-widetilde-b}
\eeqa
\end{subequations}
where the last bound follows from \eqref{eq:Cherenkov-bound-bsquare}
with $\widetilde{\sigma}_4\;\widetilde{F}=3 \times 10^{-23}$ inserted.
But even the ``weakened'' bound \eqref{eq:Cherenkov-bounds-concl-widetilde-b}
is already quite remarkable compared to the previous bound
\eqref{eq:LEP-bound} mentioned in the Introduction and, more generally, to
what can be achieved with particle accelerators on Earth (the upcoming
Large Hadron Collider at CERN will have a proton beam energy
$E_{p} \approx 7 \times\, 10^{3}\,\mathsf{GeV}$).
In fact, bound \eqref{eq:Cherenkov-bounds-concl-widetilde-b}
appears to rule out so-called $\mathsf{TeV}$--gravity models;
see Endnote [49] of Ref.~\cite{BernadotteKlinkhamer2007}, which
contains further references.

Still, the most interesting bound may very well be
\eqref{eq:Cherenkov-bounds-concl-F}, as it rules out
a single-scale classical spacetime foam having
approximately equal effective defect size $\widetilde{b}$
and separation $\widetilde{l}$, with an excluded-volume factor
$\widetilde{F} \equiv (\widetilde{b}/\widetilde{l})^4 \sim 1$.
Moreover, this conclusion  holds for arbitrarily small values of the
effective defect size $\widetilde{b}\,$, as long as a classical \st~makes sense.

More generally, also the nineteen Lorentz-violating deformation parameters
of the \modMaxth~\eqref{eq:model2-action} are strongly
bounded, as discussed in Sec.~\ref{sec:UHECRboundsModel2}:
\beq\label{eq:kappa-bound}
0 \;\leq\; |\kappa^{\mu\nu\rho\sigma}| \;\lesssim\; 10^{-18} \,,
\eeq
where, for the sake of argument, nine ``one-sided'' Cherenkov bounds
from \eqref{eq:vecalpha-Cherenkov-bound} have
been made ``two-sided'' (the even stronger astrophysics bounds
\cite{KosteleckyMewes2002} on the ten birefringent parameters are already
two-sided).
A similar bound has been obtained for the stochastic model mentioned a few
lines below \eqref{eq:alpha-parameters}; see, in particular, bound (11b) from
Ref.~[17(b)].   %%FRK--ref by hand
%%Ref.~\cite[(b)]{KlinkhamerRupp2004}.

\emph{A priori}, one expects O(1)
effects for \eqref{eq:Cherenkov-bounds-concl-F} and  \eqref{eq:kappa-bound} if
there is some kind of nontrivial small-scale structure of classical spacetime.
This expectation is based on an early observation by
Veltman \cite{Veltman1981} that, if a symmetry (for him, gauge invariance)
of the quantum field theory considered is violated by the high-energy
cutoff $\Lambda$ (or by a more fundamental theory),  then, without fine tuning,
the low-energy effective theory may contain symmetry-violating terms
which are not suppressed by inverse powers of the cutoff energy
$\Lambda$. The same observation holds for Lorentz invariance, as emphasized
recently by the authors of Ref.~\cite{Collins-etal2004}.

Contrary to these expectations, the experimental bounds
\eqref{eq:Cherenkov-bounds-concl-F} and  \eqref{eq:kappa-bound}
seem to indicate that Lorentz invariance remains perfectly
valid down towards smaller and smaller distances. In fact, this conclusion
would appear to hold down to distances at which the classical--quantum
transition of \st~occurs (generally considered to be equal to the so-called
Planck length).

In the next section, we will try to make a first step towards
understanding the robustness of Lorentz invariance,
starting from a physical picture based on a dynamic spacetime and
quantum uncertainty (entirely in the spirit of Einstein and Heisenberg).
It may be of interest to mention one other possible explanation
in the context of so-called ``emerging-gravity'' models
\cite{Bjorken2001,Laughlin2003,FroggattNielsen2005,Volovik2007},
which relies on the existence of a trans-Planckian energy scale
in a fundamental Lorentz-violating theory
with a topologically protected Fermi point \cite{KlinkhamerVolovik2005}.
But this avenue will not be pursued further here.

%%\newpage%%FRK
\section{QUANTUM-SPACETIME CONJECTURES}
\label{sec:CONJECTURES}

The main conclusion from Sec.~\ref{sec:UHECRboundsImplications}
is that Lorentz invariance appears to remain valid down towards smaller and
smaller distances, even down to distances at which the classical--quantum
transition of \st~is believed to occur. That distance is usually taken to
be the Planck length \cite{Wheeler1968,Planck1899,Wilczek2005},
\beq
 l_\mathsf{Planck} \equiv \sqrt{\hbar\,G/c^3}
  \approx 1.6 \times 10^{-35}\,\mathsf{m}
  \approx \hbar\, c/ \big( 1.2\times 10^{19}\,\mathsf{GeV} \big)\,.
\label{eq:lPlanck}
\eeq
The question, however, is whether or not quantum \st~effects show up
\emph{only} at distances of the order of the Planck length. Perhaps a quantum
\stf~\cite{Wheeler1957,Wheeler1968,Hawking1978,Visser1996}
could arise primarily from gravitational self-interactions which need
not involve Newton's constant $G$ describing the gravitational coupling
of matter (similar to the case of a gas of in\-stan\-tons in \YMth).

This brings us to the following suggestion:
\newtheorem{guess}{Conjecture}
\begin{guess}
Quantum \st~has a fundamental length scale $l$ which is
essentially different from the length $l_\mathsf{Planck}$
as defined by \eqref{eq:lPlanck}.
\label{th:Conjecture1}
\end{guess}
Concretely, this length $l$ could have a different dependence on $G$
than the Planck length. In the next two subsections, we will argue that the
standard theory appears to leave some room for a new fundamental
constant $l$ and, in the third subsection, we will present a further
conjecture as to what physics may be involved.

%%\newpage%%FRK
\subsection{Generalized Action}
\label{sec:ConjecturesGeneralizedAction}

Our starting point is Feynman's insight \cite{Feynman1948} that
the quantum world of probability amplitudes for matter is governed
by the complex phase factor $\exp(\mathrm{i}\,\mathcal{I}_\mathsf{M})$
with a real phase $\mathcal{I}_\mathsf{M}=S_\mathsf{M}/\hbar$ expressed
in terms of the classical matter action $S_\mathsf{M}$
and the reduced Planck constant $\hbar \equiv h/2\pi$.

As the merging of quantum mechanics and gravitation is far from understood,
it may be sensible to consider a \emph{generalized}
dimensionless action for ``quantum gravity''
(or, better, ``quantum spacetime'') as probed by classical matter:
\beq
\mathcal{I}_\mathsf{\,grav} = \frac{-1}{16 \pi \,l^2} \int
\d^4 x \sqrt{|g(x)|}\:\big( R(x)+2\lambda \big) + \frac{G/c^3}{l^2} \int
\d^4 x \sqrt{|g(x)|}\: \mathcal{L}_\mathsf{M}^\mathsf{\,class}(x)\,,
\label{eq:Sgeneralized}
\eeq
with the Ricci curvature scalar $R(x)$,
a new fundamental length scale $l$,
and a nonnegative cosmological constant $\lambda \geq 0$.

Two remarks may be helpful. First, expression \eqref{eq:Sgeneralized} is
to be used only for a rough description of possible quantum
effects of \st, not for those of the matter fields which are considered to
act as a classical source with coupling constant $G$.
Second, the special case $l^2 = l_\mathsf{Planck}^2$
reduces the dimensionless action \eqref{eq:Sgeneralized} to the
standard expression \cite{Visser1996},
with the Einstein-Hilbert integral multiplied
by the factor $-c^3/(16 \pi \,G)$ and the matter integral by $1/\hbar$.
Here, however, we wish to explore the possibility that $l$ is
an entirely new length scale independent of $l_\mathsf{Planck}$
(Conjecture~\ref{th:Conjecture1}).

The overall factor $l^{-2}$ in \eqref{eq:Sgeneralized}
is irrelevant for obtaining the classical field equations,
\beq
R^{\mu\nu}(x) - \half\, g^{\mu\nu}(x)\, R(x) - \lambda\,g^{\mu\nu}(x) = -
8\pi \, G \, T_\mathsf{M}^{\mu\nu}(x)\;,
\label{eq:classical-field-equations}
\eeq
where the matter energy-momentum tensor
$T_\mathsf{M}^{\mu\nu}(x)$ is defined by an appropriate functional derivative of
$\textstyle{\int}\d^4 x\,\sqrt{|g|}\;\mathcal{L}_\mathsf{M}^\mathsf{\,class}$.
But, following Feynman, the physics of a genuine quantum spacetime
would be governed by the complex phase factor
$\exp(\mathrm{i}\,\mathcal{I}_\mathsf{\,grav})$ and the overall factor
of $l^{-2}$ in \eqref{eq:Sgeneralized} would be physically relevant.
Just to be clear, we do not claim that a future theory of
quantum gravity must necessarily be formulated as a path integral but only
wish to use the Feynman phase factor as a heuristic device in order
to say something about the typical length scale involved.

The expression \eqref{eq:Sgeneralized} for the generalized
quantum phase also suggests that, as far
as \st~is concerned, the role of Planck's constant $\hbar$ would be
replaced by the squared length $l^2$, which might loosely be called the
``quantum of area'' (see, e.g., Ref.~\cite{RovelliSmolin1995} for a
calculation in the context of loop quantum gravity and
Refs.~\cite{Mead1964,Garay1995} for general remarks on a possible minimum length).
Planck's constant $\hbar$ would continue to play a role in the description
of the matter quantum fields.

However, with $\hbar$ and $l^2$ being logically independent, it is possible to
consider the ``limit'' $\hbar \to 0$ (matter behaving classically)
while keeping $l^2$ fixed (spacetime behaving nonclassically).
Even if the numerical values of the lengths $l$
and $l_\mathsf{Planck}$ will turn out to be close (or equal) in the end,
it may be conceptually interesting to consider phases of the theory with
ratios $l/l_\mathsf{Planck}$ very different from unity.

%%\newpage%%FRK
\subsection{Fundamental Constants}
\label{sec:ConjecturesFundamentalConstants}

The fundamental dimensionful constants encountered in the previous
subsection are summarized in Table~\ref{tab:fund-constants} on the
next page. In this contribution, we mainly consider
the second and third columns of Table~\ref{tab:fund-constants}
(removing the parentheses around $G$) and leave
the rigorous treatment of \emph{all} columns to a future theory.
In this subsection, however, we already make some speculative remarks
on a possible new theory encompassing all of the physics arenas
appearing in Table~\ref{tab:fund-constants}.

In that future theory, ``classical gravitation'' may perhaps
be induced \cite{Sakharov1967Visser2002}
by \emph{combined} quantum effects of matter and spacetime, giving
Newton's gravitational constant
\beq
 G=\zeta\,c^3\,l^2/\hbar\:,
\label{eq:G}
\eeq
with a calculable numerical coefficient $\zeta\geq 0$.\footnote{There may
or may not be a connection with the ``emergent-gravity'' models mentioned
in the last paragraph of Sec.~\ref{sec:UHECRboundsImplications}. See also
Ref.~\cite{Volovik2006} for an interesting comparison to hydrodynamics.}
In this way, the ``large'' classical coupling constant $G$
would be the ratio of ``small'' quantum constants
$l^2$ and $\hbar$. Practically, the numerical coefficient $\zeta$ gives
the ratio $l_\mathsf{Planck}^2/l^2$.

The dimensionful constants $\hbar$, $c$, and $l^2$ of
Table~\ref{tab:fund-constants} correspond to, respectively,
the quantum of action, the limiting speed of a massive particle,
and the quantum of area (up to an as yet undetermined numerical factor).
In addition, each of
these constants is believed to play ``the role of a conversion factor
that is essential to implementing a profound physical concept''
(quoting from Ref.~\cite{Wilczek2005}).

\begin{table}[t]
\begin{tabular}{ccc}
\hline
    \tablehead{1}{c}{b}{quantum matter}
  & \tablehead{1}{c}{b}{classical relativity}
  & \tablehead{1}{c}{b}{quantum spacetime}\\
\hline
$\hbar$ & $c\;(G)$ & $l^2$\\
\hline
\end{tabular}
\caption{Fundamental dimensionful constants of nature,
including the hypothetical quantum of area $l^2$.
The gravitational constant $G$ appears between brackets,
as it can, in principle, be expressed in terms of the other
constants; see \eqref{eq:G} in the main text.}
\label{tab:fund-constants}
\end{table}

\begin{figure}[b]
\label{fig:fund-constants-triangle}
\thicklines
\vspace*{-2cm}
\hspace*{-5.5cm}
\begin{picture}(100,100)(0,0)
\put(45,50) {$\mathsf{LENGTH}$}
\put(100,60){\vector(4,1){80}}\put(118,80) {$M[\,1/c\,]$}
\put(190,80){$\mathsf{TIME}$}
\put(203,73){\vector(0,-1){38}}\put(210,50) {$M[\,\hbar/c^2\,] \circ I$}
\put(190,20){$\mathsf{MASS}$}
\qbezier[25](180,25)(140,35)(100,45)\put(98,45.5){\vector(-4,1){0}}
\put(113,15) {$M[\,l^2\,c/\hbar\,]$}
 \vspace*{-1cm}
\end{picture}
\caption{Fundamental dimensionful constants $\hbar$, $c$, and $l^2$
from Table~\ref{tab:fund-constants}
in the role of conversion factors, with numerical coefficients omitted.
The operation $M[\,z\,]$ stands for multiplication by $z$
and the operation $I$ for inversion.
Hence, time is first inverted (to give a frequency)
and then multiplied by $\hbar/c^2$ (to give a mass).
The linear relation between mass and length is discussed in
Sec.~\ref{sec:ConjecturesFundamentalConstants} of the main text.
\vspace*{-0cm}}
\end{figure}
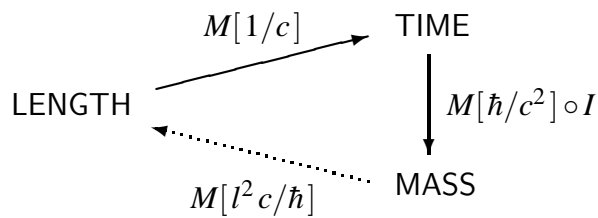

The role of the individual constants $c$ and $\hbar$ is clear:
the first converts time into length according to relativity theory
($c\,T=L$, where physical quantities to be converted are denoted by
capital letters) and the second converts frequency or inverse time
into energy according to quantum theory ($\hbar\:\Omega=\hbar\:2\pi/T=E$).
From Einstein's theory of gravity and relation \eqref{eq:G},
it would seem that the combination $l^2/(\hbar\, c)$ acts as
the conversion factor between space curvature
(with dimensions of inverse length square)
and energy density
(with dimensions of energy over length cube),
so that $1/L^2 \sim \zeta\,l^2/(\hbar\, c)\,E/L^3$, with the numerical
coefficient $\zeta$ from \eqref{eq:G}.
Replacing the energy $E$ by $M\,c^2$, these conversion factors
are summarized in Fig.~\ref{fig:fund-constants-triangle}.
Starting with $\mathsf{LENGTH}$ on the left of this figure and going
around clockwise, the required number of fundamental constants is seen
to accumulate steadily.

In the previous paragraph, we have found the
conversion factor between mass and length by relying
on Einstein's theory of gravity. Incidentally, the simplest way to
get this conversion factor may be to use the standard result for
the Schwarzschild radius of a central mass $M$, $R_\mathsf{Schw}\equiv 2 GM/c^2$,
and to replace $G$ therein by \eqref{eq:G}.
For the corresponding arrow in Fig.~\ref{fig:fund-constants-triangle},
this would imply that the conversion factor $l^2\,c/\hbar$ comes
together with the pure number $\zeta$ appearing in \eqref{eq:G}.
However, it might be interesting, as discussed
in the last paragraph of Sec.~\ref{sec:ConjecturesGeneralizedAction},
to consider a possible phase of the theory with $\zeta=0$ and still keep
the linear relation between mass and length.

For $\zeta=0$, the physical interpretation of the
conversion factor $l^2\,c/\hbar$ in Fig.~\ref{fig:fund-constants-triangle}
is entirely open (hence, the dotted arrow in the figure).
A trivial suggestion (definitely not yet a ``profound physical concept'')
would run as follows.
Given the fundamental length scale $l$ and the constants
$c$ and $\hbar$, a corresponding energy density is
defined as $\widetilde{\rho}\equiv \hbar\, c/l^4$.
Then, one possible physical interpretation would be that the mass $M$
of an ``elementary particle'' is associated to a characteristic length
$\widetilde{L}$ in such a way that
the fundamental energy density $\widetilde{\rho}$ over a volume
$\widetilde{V}$
obtained by multiplying $\widetilde{L}$ with the minimum area $l^2$,
$\widetilde{V}\equiv \widetilde{L}\:l^2$,
returns the original mass, $M\equiv \widetilde{V}\, \widetilde{\rho}$.
In this way, the associated length $\widetilde{L}$ of a mass $M$ would
correspond  to the largest linear spread a fiducial
volume $V\equiv M/\widetilde{\rho}$
could have, provided the area of any cross section of the volume $V$
is not allowed to drop under the value $l^2$. The heuristic idea
behind this length $\widetilde{L}$ is that an inserted material
mass $M$ distorts the intrinsic foam-like structure of space.

Observe that the mass dependence of
the associated length $\widetilde{L}\equiv M\,l^2\,c/\hbar$
is opposite to that of the standard
Compton (reduced) wavelength $\lambdabar_\mathsf{C}\equiv \hbar/(M\,c)$.
For the known elementary particles, with $M$ of the
order of  $10^2\:\mathsf{GeV}/c^2$ or less, and $l$ below the value
quoted in \eqref{eq:Cherenkov-bounds-concl-widetilde-b},
the mass-induced fuzziness scale $\widetilde{L}$ of a particle will
be very much smaller than
the corresponding Compton wavelength $\lambdabar_\mathsf{C}$.
Specifically, the ratio of these two length scales is given by
the inverse ratio of the corresponding squared energies,
$\widetilde{L}/\lambdabar_\mathsf{C}$ $=$ $\big(M\,c^2/(\hbar\, c/l)\big)^2$.

At this moment, it may be helpful to list all the length scales
possibly relevant to an elementary particle with mass $M$
(in order of increasing size,
taking $\zeta < 1/2$ and temporarily setting $c=\hbar=1$):
the Schwarzschild radius $R_\mathsf{Schw}\equiv 2\,\zeta\,l^2\,M$,
the mass-induced fuzziness scale $\widetilde{L}\equiv l^2\,M$,
the fundamental quantum-spacetime length scale $l$,
and the Compton wavelength $\lambdabar_\mathsf{C}\equiv 1/M$.
In principle, the mass-induced fuzziness scale $\widetilde{L}$
(or perhaps even the larger intrinsic fuzziness scale $l$) could
lead to form-factor effects in high-energy scattering processes and
possibly affect the renormalization of the standard-model quantum field theory.
A minimum length may even play a crucial role in
obtaining the usual rules of quantum mechanics,
as discussed in Ref.~\cite{BuniyHsuZee2006}.

%%\newpage%%FRK
\subsection{Vacuum Energy Density}
\label{sec:ConjecturesVacuumEnergyDensity}

Let us return to the generalized action \eqref{eq:Sgeneralized},
with Newton's gravitational constant $G$ appearing explicitly
but without Planck's constant $\hbar$.
If Conjecture~\ref{th:Conjecture1} holds true, the question
arises as to what type of physics determines the length scale $l$.
Here, we assume this physics to be independent of $\hbar$, otherwise the
discussion would be more along the lines sketched in the previous subsection.
One possible answer would then be given by the following suggestion:
\begin{guess}
The quantum-spacetime length scale $l$ is related to a
nonvanishing cosmological constant or vacuum energy density.
\label{th:Conjecture2}
\end{guess}

For the case of the early universe,
with a vacuum energy density $\rho_\mathsf{vac} \equiv
E_\mathsf{vac}^4/(\hbar\,c)^3$
in the matter part of the action \eqref{eq:Sgeneralized}
and $\lambda=0$ in the geometric part, it can be argued \cite{Klinkhamer2007}
that the following approximate relation holds
($\hbar$ appears only as an auxiliary constant):
\beq
 l
 \,\stackrel{?}{\sim}\,
 c^2/\sqrt{G\,\rho_\mathsf{vac}}
 =
 \hbar\,c\;E_\mathsf{Planck}/E_\mathsf{vac}^2
 \approx
 2 \times 10^{-29}\,\mathsf{m}\,
 \left( \frac{E_\mathsf{Planck}}{10^{19}\,\mathsf{GeV}}\right)
 \left( \frac{10^{16}\,\mathsf{GeV}}{E_\mathsf{vac}} \right)^2,
\label{eq:l-conjecture-numerical}
\eeq
where the Planck energy scale is given by $E_\mathsf{Planck}$ $\equiv$
$\hbar\,c/\lP$
and the numerical value for $E_\mathsf{vac}$ has been identified with the
``grand-unification'' scale suggested by elementary particle physics
\cite{ItzyksonZuber1980,GeorgiQuinnWeinberg1974}.
As anticipated in the sentence under Conjecture~\ref{th:Conjecture1},
the length scale $l$ as given by the first mathematical expression
on the right-hand side of \eqref{eq:l-conjecture-numerical}
has a $G$ dependence which is different from that of the standard
Planck length \eqref{eq:lPlanck}.

If \eqref{eq:l-conjecture-numerical} holds
true with $\lP / l \sim 10^{-6}$, it is perhaps possible to have
\emph{sufficiently rare} defects left over from the crystallization
process of an initial quantum \stf~to a classical \st.\footnote{It is
likely that a proper understanding of this crystallization process requires
new ideas about the quantum measurement problem and possibly even
a mechanism for objective state reduction \cite{Penrose2005}.}
With the effective defect size
$\widetilde{b}$ set by $\lP$ (matter related) and the effective defect
separation $\widetilde{l}$ set by $l$ (vacuum related), these spacetime
defects would give the following excluded-volume factor in the modified
photon dispersion relation \eqref{eq:disprel-general-form-photon}:
\beq \widetilde{F}  \equiv
\big(\:\widetilde{b}/\widetilde{l}\;\big)^4 \,\stackrel{?}{\sim}\, 10^{-24},
\eeq
which is close to saturating the current UHECR bound
\eqref{eq:Cherenkov-bound-ratio}.
As mentioned at the end of Sec.~\ref{sec:UHECRboundsModel1},
a possible detection of Lorentz-violating effects
(having ruled out conventional explanations)
would be an entirely different matter than setting better and better bounds
on the small-scale structure of spacetime,
even though the latter type of ``null experiments'' can also be of great
importance (think of the role of the Michelson--Morley experiment
for the discovery of special relativity \cite{Pais1982}).

%%\newpage%%FRK
\section{SUMMARY}
\label{sec:SUMMARY}

The main phenomenological conclusion of the first
part of this review (Secs.~\ref{sec:PHENOMENOLOGY}
and \ref{sec:UHECR-CHERENKOV-BOUNDS})
is that quantum \stf, if at all real, appears to have given rise
to a classical spacetime manifold which is remarkably smooth.
Using astrophysics data, this result can be quantified as follows:
the defect excluded-volume factor in the
modified proton and photon dispersion relations
(\ref{eq:disprel-general-form}ab)
is bounded by $\widetilde{F} \lesssim  10^{-23}  \ll 1$
and the Lorentz-violating parameters
in the \modMaxth~\eqref{eq:model2-action} by
$|\kappa^{\mu\nu\rho\sigma}| \lesssim 10^{-18} \ll 1$.

Prompted by this phenomenological conclusion,
a theoretical suggestion has been advanced
in the second part (Sec.~\ref{sec:CONJECTURES}),
namely, that the quantum theory of spacetime may
have a fundamental length scale $l$ conceptually different from
the Planck length (Conjecture~\ref{th:Conjecture1})
and possibly related to vacuum energy density
(Conjecture~\ref{th:Conjecture2}).
With two length scales present, $l$ and $l_\mathsf{Planck}$,
it is perhaps possible to
satisfy the tight experimental constraint on $\widetilde{F}$.

Still, it is safe to say that the quantum origin of
classical \st~remains a mystery.

\begin{theacknowledgments}
It is a pleasure to thank the organizers for bringing about
this interesting ``Symposium on Gravitation and Cosmology''
(El Colegio Nacional, Mexico City, September 2007)
and the participants for informative discussions. The participants of
another meeting, ``Condensed Matter Meets Gravity''
(Lorentz Center, Leiden, August 2007),
are also thanked for equally informative discussions.
\end{theacknowledgments}

%%\newpage%%FRK

\end{document}